\documentclass[aps,prd,twocolumn,amsmath,amssymb,amsfont,superscriptaddress,nofootinbib]{revtex4-1}
\usepackage{amsmath,aas_macros,xcolor,graphicx}
\newcommand{\bs}{\boldsymbol}
\newcommand{\diff}{{\mathrm d}}

\newcommand{\T}{\bold{T}}
\newcommand{\I}{\bold{I}}
\newcommand{\spin}{\bs{j}}

\begin{document}


\title{Parity-odd neutrino torque detection}


\author{Hao-Ran~Yu}\email{\url{haoran@cita.utoronto.ca}}
\affiliation{Tsung-Dao Lee Institute, Shanghai Jiao Tong University, Shanghai, 200240, China}
\affiliation{Canadian Institute for Theoretical Astrophysics, University of Toronto, M5S 3H8, Ontario, Canada}
\affiliation{Department of Astronomy, Shanghai Jiao Tong University, Shanghai, 200240, China}

\author{Ue-Li~Pen}\email{\url{pen@cita.utoronto.ca}}
\affiliation{Canadian Institute for Theoretical Astrophysics, University of Toronto, M5S 3H8, Ontario, Canada}
\affiliation{Tsung-Dao Lee Institute, Shanghai Jiao Tong University, Shanghai, 200240, China}
\affiliation{Dunlap Institute for Astronomy and Astrophysics, University of Toronto, M5S 3H4, Ontario, Canada}
\affiliation{Canadian Institute for Advanced Research, CIFAR Program in Gravitation and Cosmology, Toronto, M5G 1Z8, Ontario, Canada}
\affiliation{Perimeter Institute for Theoretical Physics, Waterloo, N2L 2Y5, Ontario, Canada}

\author{Xin~Wang}\email{\url{xwang@cita.utoronto.ca}}
\affiliation{Canadian Institute for Theoretical Astrophysics, University of Toronto, M5S 3H8, Ontario, Canada}


\date{\today}

\begin{abstract}
Cosmological observations are promising ways to improve our understanding of neutrino mass properties. The upper bound on the sum of masses is given by the cosmic microwave background and large scale structure. These measurements are all parity-even, and potentially contaminated by unmodeled baryonic effects. In this paper we propose a novel parity-odd gravitational effect of neutrinos: A unique contribution to the directions of the angular momentum field of galaxies and halos. This observable is free of contamination in linear perturbation theory, and thus likely more cleanly separated from other nongravitational effects. A deep 21-cm survey to redshift 1 can potentially yield a $5\sigma$ significance on neutrino mass detection for a fiducial sum of neutrino masses of 0.05 eV.
\end{abstract}


\maketitle

\section{Introduction} 
Neutrino mass is a long-standing physics problem. 
The flavor oscillation experiments \citep{2002PhRvL..89a1301A} discovered the mass 
splittings of neutrinos and placed a lower bound of the sum of their mass 
$M_\nu \equiv \sum_{i=1}^3 m_{\nu_i} \gtrsim$ 0.05 eV \citep{2014ChPhC..38i0001O}. 
The existence of neutrino mass has profound impacts on cosmic evolution, and the current cosmic microwave background (CMB) observations provide an 
upper bound of $M_\nu\lesssim 0.12$ eV \citep{2018arXiv180706209P}. 
At low redshifts, neutrinos become nonrelativistic, 
and contribute to the matter energy density $\Omega_m$ in the structure formation. 
Unlike the majority of matter, the cold dark matter (CDM) and baryons, 
neutrinos maintain a high velocity dispersion, 
referred to as ``free-streaming,'' which reduces their gravitational collapse on small scales. 
A number of large scale structure (LSS) surveys \citep{2011arXiv1110.3193L,2015AAS...22533605E}
plan to improve this upper bound using neutrino effects on LSS. Their approaches either measure the impact of neutrinos on the growth of structure by measurement at different redshifts, or on different length scales.

The small gravitational contribution from neutrinos requires a precise measurement at two different epochs or scales. Most techniques aim to combine the CMB with a second, lower redshift, and use small scales where plenty of modes are available for a precise measurement. To measure the impact on scale dependence requires the measurement on a larger scale, where cosmic variance limits the number of observable modes. Here we propose to exploit a historic {\it fossil} comparison at large scales: we measure the {\it same} modes at two redshifts in the full 3-D LSS volume, which overcomes the cosmic variance limit. Neutrinos free stream, and are mostly smooth below their free streaming scale, tens of Mpc. Baryons are conserved, and even in the most extreme feedback scenarios do not move more than 10 Mpc relative to CDM. In this paper we present a novel approach to exploit the torquing effect by neutrinos to probe their mass. This is meant as a first exploration, intended to stimulate observational and more detailed theoretical studies. 


\section{Spin of galaxies}\label{sec.spin}
The {\it direction} of angular momentum (hereafter {\it spin}) of a galaxy is readily observable, while the magnitude is not. 
The majority of galaxies are disk galaxies, and the rotation axis is perpendicular to the disk \citep{1998MNRAS.297L..71M}. 
The orientation of the disk is obtained by inclination and parallactic angles (parity-even) and dust absorption, and from the Doppler effect of spectral lines we determine the plus-minus sign of the spin, which is parity-odd.
A parity-odd spin measurement cannot be contaminated by the linear perturbation theory. 
Only gravity affects spin statistics of halos or galaxies at large separations, since baryons do not travel to the distance of the free-streaming of neutrinos.

We present the neutrino torque effect, to modulate the spin of dark matter halos.
Here we use halos to represent galaxies.
The galaxy spin can be observed to a much larger radii through the 21-cm radio emission of cold gas, and the spin change only very modestly to larger radii \citep{2001ARA&A..39..137S}. 
Hydrodynamic galaxy formation simulations suggest that spin directions of stellar, 
cold gas components, and CDM halo are correlated (e.g. \citep{2010MNRAS.404.1137B,2010MNRAS.405..274H,2011MNRAS.415.2607D,2015ApJ...812...29T} and references therein)\footnote{We
note that although the angular momentum \emph{magnitudes} of dark matter halos and
baryons are not correlated, their angular momentum \emph{directions} (spins) are 
highly correlated (e.g. section 4.2 of \citep{2018arXiv180407306J}), with a median
misalignment angle typically $30^\circ$.}.
The magnitudes are not as correlated \citep{2018arXiv180407306J}, as one might expect from the
substantial dissipation and outflow.
The observed mostly coplanar rotation of baryons in late type galaxies indicates that additional dissipation or selective outflow are unlikely to change the angular momentum direction,
which is common to most baryons, while the magnitude is likely to change, and varies within the baryonic inventory.

\section{Theory}\label{sec.theory}
In the picture of LSS formation, gravitational instability lets initial density fluctuations form dark matter halos, where galaxies are embedded.
In these highly nonlinear structures, uncertainties of halo bias, 
halo merging history and baryonic mechanisms obstruct us from clearly 
understanding the statistics like number counts and morphologies.
In comparison, the spins of galaxies/halos represent a local probe of gravity, 
especially contributed from the linear epoch of the structure formation. 

The initial halo spin is written in Lagrangian space as 
$\spin_L\propto-\int_{V_L}\bs{q}\times\bs{\nabla}\phi_c\diff^3\bs{q}$, 
where $\phi_c$ is the gravitational potential of CDM, 
$\bs{q}$ is the Lagrangian coordinates relative to the center of mass of the protohalo in volume $V_L$. 
In the tidal torque theory \citep{1984ApJ...286...38W},
it can be written as,
$\spin_T\propto\bs{\epsilon}\,\I_q\,\T_c$, where
$\I_q=(I_{ij})\equiv(\int_{V_L} q_i q_j \diff^3\bs{q})$,
$\T_c=(T_{ij})\equiv(\partial_i\partial_j\phi_c)$
are the protogalactic inertia tensor and the local tidal shear tensor\footnote{$\I_q$ 
and $\T_c$ differ from textbook by trace, which does not contribute to $\spin_T$.}, 
and $\bs{\epsilon}=(\epsilon_{ijk})$ is the Levi-Civita symbol to collect the asymmetric components generated by
the misalignment between $\I_q$ and $\T_c$. Here, $\I_q$ and $\T_c$ are parity-even, $\bs{\epsilon}$ is parity-odd, and thus $\spin_T$ is parity-odd.
In linear perturbation theory, $\phi_c$ remains constant, so $\T_c$ decays as $a^{-2}$ due to the cosmic expansion, where $a$ is the scale factor.
Also, the protohalo shrinks in size and turns more spherical in shape.
A decayed tidal field is hard to torque small and round objects, so the spin are expected to be contributed mostly in the linear regime \citep{2002MNRAS.332..325P}.
If a halo has a merging history, it simply collects disconnected regions in Lagrangian space but conserves their total spin.
These concepts can be straightforwardly tested in $N$-body simulations.

In Lagrangian space, CDM and baryonic matter are torqued by the same gravitational shear, so 
their subsequent local evolution should not systematically change their respective spin directions.
For example, baryonic feedback has to conserve angular momentum, unless the baryons are expelled from the galaxy with systematic misaligned angular momentum.
The dynamical friction between baryons and dark matter can only increase the angular momentum correlation.
These all indicate a high correlation between CDM and baryons (e.g. \citep{2010MNRAS.404.1137B,2010MNRAS.405..274H,2011MNRAS.415.2607D,2015ApJ...812...29T,2018arXiv180407306J}).
 
Massive neutrinos contribute a subpercent fraction of the matter ingredient, and their unique spatial distribution and evolution should contribute a unique torque to CDM and baryons.
In particular, neutrino density field traces CDM on large scales while their small scale structures are smoothed out by their free-streaming. 
They contribute a predictable tidal tensor $\T_\nu(m_{\nu_i})$ depending on their mass \citep{2015PhRvD..92b3502I}. 
The interplay between $\I_q$ and $\T_\nu$ leads to an additional neutrino torque $\spin^T_\nu\propto\bs{\epsilon}\,\I_q\,\T_\nu$.
Integrated it over the cosmic evolution, we denote the total neutrino modulation of the halo spin as $\spin^T_{\nu 0}$ given by the tidal torque theory.
This torque is unique in that, the neutrino free-streaming scale is much larger than the scale to which local nonlinear and baryonic effects can systematically contribute.

\section{Reconstruction}\label{sec.reco}
The feasibility of predicting neutrino torque relies on the precision reconstruction of $\I_q$ and $\T_\nu$.

Neutrino distribution shares the same Fourier phases of CDM from halo reconstruction \citep{2015PhRvD..92b3502I,2017ApJ...847..110Y}, but only differs by the ratio of their linear transfer function, depending on the neutrino masses. The reconstruction $\T_\nu$ is reliable even if it is applied directly on halos \citep{2015PhRvD..92b3502I}, because they are relatively linear even at present epoch.

The reconstruction of $\I_q$ relies on the fact that $\I_q$ and $\T_c$ are highly correlated in Lagrangian space \citep{2000ApJ...532L...5L,2001ApJ...555..106L},
which can be understood that $\I_q$ is a collection of matter to be shell-crossed to form a halo, being parallel with $\T_c$.
The latter depends on a precise reconstruction of CDM initial conditions. 
Recent years, many emerging reconstruction methods have achieved unprecedented accuracies.
For example, the isobaric halo reconstruction \citep{2017PhRvD..96l3502Z,2018arXiv180706381W} can recover the initial conditions of the Universe from spatial distribution of halos.
In the case of high halo number densities the reconstructed density field is correlated with the true initial density field on scales $k\lesssim 0.7 h/{\rm Mpc}$ \citep{2017ApJ...847..110Y}, 
close to the limit ($k\lesssim 1 h/{\rm Mpc}$) of isobaric reconstruction from using direct CDM density field \citep{2017MNRAS.469.1968P}, 
or the limit of reconstruction from the true displacement field \citep{2017PhRvD..95d3501Y}. 
ELUCID simulations are able to reconstruct the full evolution history of the real local Universe \citep{2014ApJ...794...94W}. These reconstruction techniques enable us to study the tidal field from both CDM and neutrinos in an unprecedented precision, at different epochs of the cosmic evolution. We construct an equivalent inertia $\I_R$ from $\T_c$, in order to maximize the cross-correlation $\mu$ between $\bs{\epsilon}\,\I_R\T_\nu$ and $\spin^T_{\nu 0}$ (see Appendix \ref{app.I}).

The final neutrino torque can be understood as an interaction between two scales -- a small collapsing scale $\T_c$,
a large neutrino free-streaming scale $\T_\nu$,
and with an antisymmetric (parity-odd) operator $\bs{\epsilon}$ collecting the antisymmetric (parity-odd) contributions from the multiplication of these two symmetric (parity-even) tensors.

\begin{figure}
\centering
  \includegraphics[width=1\linewidth]{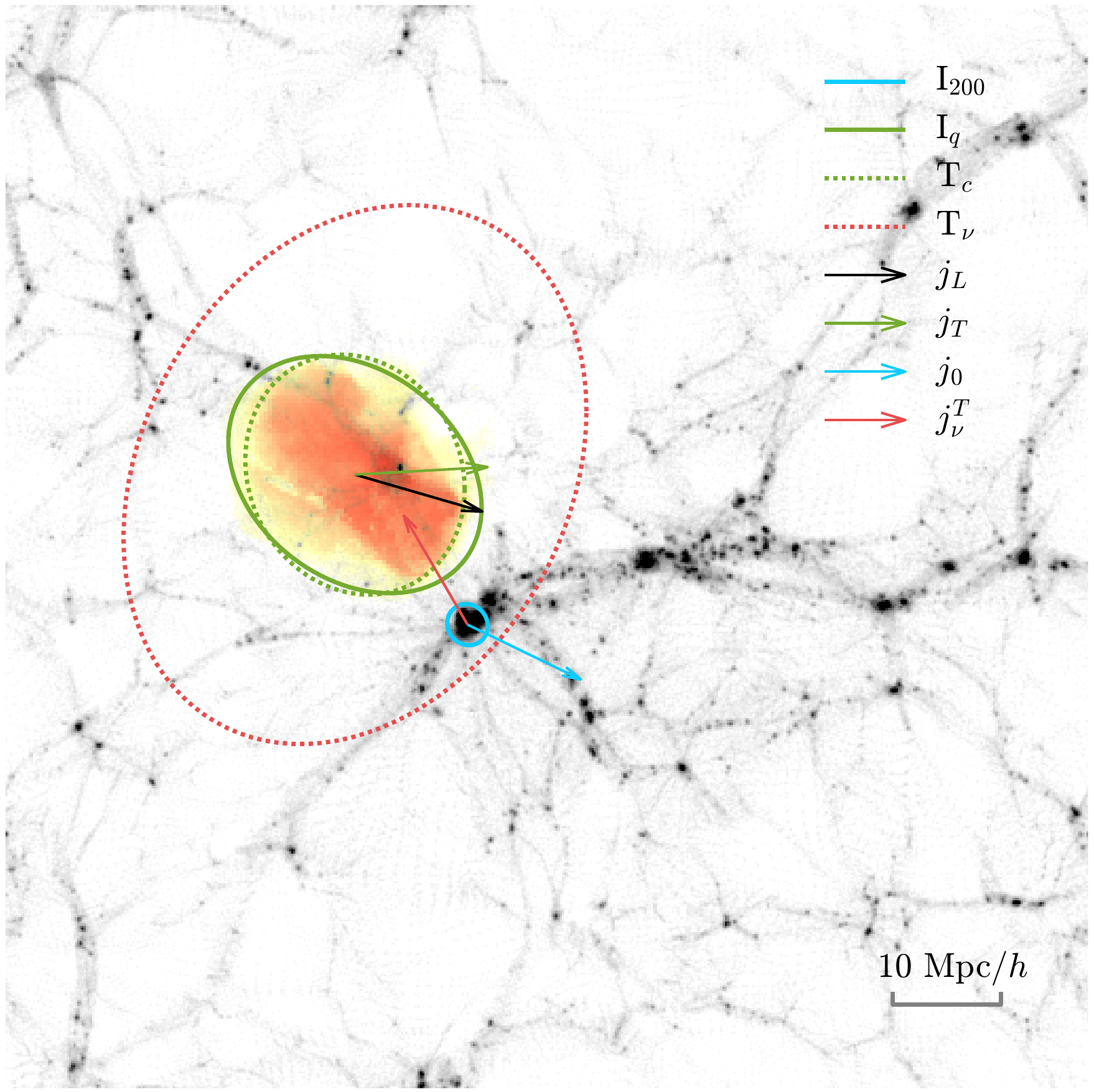}
 \caption{Visualization of the neutrino torque. 
 We show the LSS slice centered at a selected halo with depth twice the halo radius $r_{200}$. 
 Equivalent ellipsoids by solid lines show the moment of inertia in Lagrangian and Eulerian space $\I_q$ and $\I_{200}$, 
 while dotted ellipsoids show the tidal shear from CDM and neutrinos $\T_c$ and $\T_\nu$. $\spin_L$, 
 $\spin_0$ are the Lagrangian and Eulerian halo spins, whereas $\spin_T$ is the tidal torque prediction. 
 The initial neutrino torque is shown by $\spin^T_\nu$.}\label{fig.1}
\end{figure}

\section{Simulation}\label{label.sim}
These correlations and coefficients are tested across a set of high-resolution $N$-body simulations \citep{2018ApJS..237...24Y}. 
Given any halo formed in the simulation, all the belonging particles are mapped back to Lagrangian space. 
The status of this definite set of particles can be traced in a resimulation of the exact same initial conditions.

In Fig.\ref{fig.1} (all quantities are projected onto the plane of this paper), 
we select a very massive halo ($7.8\times 10^{14}M_\odot$) to maximize the clarity of the visualization of the halo properties.
We confirm in simulation that these properties (cross-correlations) have only weak dependence on halo mass. 
The background LSS at redshift $z=0$ has the thickness $2r_{200}$ with $r_{200}$ being the halo radius within which the mean halo density is 200 times the mean matter density of the Universe.
The Lagrangian mapping of this halo is shown by the protohalo's column density with the orange clouds.
To visualize the tidal torque theory, we plot equivalent ellipsoids (solid curves) with their moment of inertia equal to $\I_q$ and 
$\I_{200}$, where $\I_{200}$ is the moment of inertia within $r_{200}$. 
The ellipsoids with dotted lines correspond to $\T_c$ and $\T_\nu$, normalized such that their volumes are $V_L$ and $8V_L$ respectively. 
As expected, $\I_q$ and $\T_c$ are aligned with their primary axes in parallel with the collapsing direction,  perpendicular to the filament containing the halo. 
Their minor misalignment yields the tidal torque $\spin_T$, which is the first order approximation of the true initial spin $\spin_L$ (all the spin arrows are normalized to have 15 Mpc$/h$). 
They are, in general, highly correlated with the spin of the final halo $\spin_0$.
In comparison, the neutrino tidal shear $\T_\nu$ torques $\I_q$ in an other less correlated direction $\spin^T_\nu$.

The validity of the tidal torque formulation is tested by an ensemble average over all halos, 
across 3 orders of magnitude in mass range ($>2\times10^{12}M_\odot$), and over simulations with different resolutions.
As the Universe evolves from the initial condition to $z=0$, the cross-correlation coefficients $\left\langle \spin(z) \cdot \spin_L \right\rangle$ and
$\left\langle \spin(z) \cdot \spin_T \right\rangle$ smoothly decrease from 1 to 0.80, and
from 0.75 to 0.69, respectively.

For neutrinos, the first-order tidal torque approximation gives a near perfect (with cross-correlation 0.99) representation of the actual neutrino torque, and it has generally $<0.2$ cross-correlates with CDM torques. This is expected in that $\T_c$ dominates locally  whereas $\T_\nu$ is contributed beyond the neutrino free-streaming scale. These two species, however, have a highly correlated contribution in structure formation. When we consider the gravitational forces that the two species exerted to the protohalo, $\bs{F}_{c/\nu}\propto\int_{V_L}\bs{\nabla}\phi_{c/\nu}\diff^3\bs{q}$, the cross-correlation between two species is as high as 0.86 .

From simulations, we estimate the magnitude of integrated neutrino torque $\left\langle|\spin^T_{\nu 0}|/|\spin_0|\right\rangle\simeq 3\times10^{-4}$. In particular, the effect given by the smoother distribution for neutrinos relative to CDM accounts 0.03, while the neutrino fraction $f_\nu=3.5\times 10^{-3}$ (for $M_\nu=0.05$ eV) and the backreaction factor from neutrinos to CDM $\sqrt{8}$ \citep{2012MNRAS.420.2551B} contribute the rest.

Measured from simulations, the cross-correlation between $\bs{\epsilon}\,\I_R\T_\nu$ and $\spin^T_{\nu 0}$ is about $\mu=0.19$ (see Appendix \ref{app.I}). We then need $5\times 10^9$ halos to have a $5\sigma$ detection (see Appendix \ref{app.error}). If we improve the reconstruction of $\I_R$ from studying the cosmic evolution history \citep{2014ApJ...794...94W}, the lower limit of required halos is $2\times 10^8$. When potential observational errors are considered, including the scatter between halo and galaxy spins, more halos are required for an neutrino torque detection. For example, a $50^\circ$ average misalignment angle between halo and galaxy spins requires additional 56\% halos needed (see Appendix \ref{app.error}).

\section{Discussion}\label{sec.discussion}
We proposed a novel approach to measure the neutrino mass that has not been considered before. 
Two independent observables are compared to infer neutrino mass:
the displacement field measures the inertia tensor on small scales, 
and the neutrino tidal field on large scales.
This is compared to the angular momentum field on the same large scales, 
thus not subject to cosmic variance.
The two fields measure the tides at two different epochs: spin is a fossil of the epoch of galaxy formation.
The applicability relies on the new research studies on the angular momentum connection.

From past and current $N$-body simulations, the spin correlation between halos and their Lagrangian protohalos  
(e.g. \citep{2000ApJ...532L...5L,2002MNRAS.332..325P,2019arXiv190401029Y} and this work) is higher than 0.6, 
meaning that the nonlinear effects at low redshifts only add less than 
factor of 2 scatter to the spin correlation.
Recent galaxies formation simulations
(e.g. \cite{2010MNRAS.404.1137B,2010MNRAS.405..274H,2011MNRAS.415.2607D,2015ApJ...812...29T,2018arXiv180407306J}) also show a $>\sim 0.6$ spin correlation between galaxies and halos,
and thus baryonic effects also contribute less than factor of 2.
These strong correlations are much higher than the $\sim 0.01$ order correlation between galaxy shapes
and true tidal fields by weak lensing observations \citep{2013JCAP...01..026A}. 
We do not propose to use galaxies in clusters, and most spiral galaxies are not in clusters.

We found that the cross-correlation coefficients between $\spin_0$ and initial values $\spin_L$ and $\spin_T$ are prominently higher than our previous understandings (e.g. \citep{2000ApJ...532L...5L}). 
We carefully investigate the numerical errors that may affect the results. 
P3M (particle-particle particle-mesh) algorithms result in higher cross-correlation between initial and final spins, compared to PM (particle-mesh), where additional tangential forces in PM violate the angular momentum conservation. 
Higher mass halos in a given simulation generally have slightly higher cross-correlations between initial and final spins, however the correlation is enhanced as we use higher mass resolutions. 
All other cross-correlation measurements have only weak dependencies on the halo mass, even in a fixed simulation. 
With different box sizes, mass resolutions, force resolutions, we find that the results are consistent across these simulations. 
The estimation of number of halo spins needed to detect the neutrino torque depends on $\mu=0.19$. This has a very weak dependence on halo mass and configuration of the simulation. An accurate study of the neutrino torque requires mass and force resolutions to cover the wide range of halo mass, a large box size ($>600\,{\rm Mpc}/h$) to account the neutrino tides at distance, and neutrino particles/fluids, further studies of reconstruction of $\I_R$ (Appendix \ref{app.I}) to calculate more precise nonlinear neutrino effects on an evolving halo. These require future simulations with computing power comparable to that of TianNu \citep{2017RAA....17...85E}.

Surveys like the Hubble Sphere Hydrogen Survey (HSHS) \citep{2006astro.ph..6104P} or a modified 21cm Cosmic Vision \citep{2018arXiv180207216D} are able to observe order of $10^9$ HI galaxies ($>10^{12}M_\odot$), 
in a cosmic volume $(4{\rm Gpc}/h)^3$ below redshift $z\simeq 1$ \citep{2004MNRAS.350.1210Z}. 
Under the standard model of the Universe with $M_\nu=0.05$ eV, a $5\sigma$ detection will be reachable.
Beyond standard models, e.g., the neutrino mass could be
generated through a neutrino vacuum condensate triggered by a gravitational $\theta$-term
\citep{2016PhRvD..93k3002D}, the tidal torque history will be different, 
and we need different numbers of galaxies to differentiate between these models.


\section{Conclusion}\label{sec.conclusion}
Enormous efforts have been contributed to the neutrino mass properties and new, independent cosmological neutrino mass effects are proposed to constrain the neutrino mass \citep{2014PhRvL.113m1301Z,2016PhRvL.116n1301Z,2017NatAs...1E.143Y,2017PhRvD..95h3518I,2019PhRvL.122d1302C}.
Radio astronomic and cosmological surveys are promising low-cost experiments implemented in the Universe to probe basic physics mysteries. 
The angular momentum is a 3-dimensional, gravity-driven, parity-odd measure, 
and is readily observable, well modeled by the tidal torque theory, and well conserved over the cosmic evolution.
It contains comparable amount of information as density field but poorly appreciated in application.
Recent developed reconstruction techniques enable us to visit an-order-of-magnitude more precise initial conditions of the Universe in Lagrangian space, 
to accurately reconstruct the neutrino torque effects,
and to unveil the neutrino mass properties in upcoming galaxy surveys.

\acknowledgments
We thank the anonymous referee who gave valuable comments which improved this paper.
H.R.Y. thanks Derek Inman, Yipeng Jing, Pengjie Zhang and Jiaxin Han for helpful comments and suggestions.
We acknowledge funding from NSERC.
The simulations were performed on the Sunnyvale cluster at CITA and on the Niagara supercomputer at the SciNet HPC Consortium.

\appendix
\section{RECONSTRUCTION OF $\I_R$}\label{app.I}
In Lagrangian space, $\I_q$ and $\T_c$ are highly correlated.
In linear, intermediate epochs, $\I_q$ is first reshaped according to the dominating $\T_c$, thus the linear evolved $\T_\nu(\tau)$ will act on an evolved $\I(\tau)$.
Further, $\I(\tau)$ will be affected by nonlinear effects, which are more difficult to predict, however halos are relatively small and their spin directions can hardly be changed dramatically.
A deep study of the latter steps above is beyond the scope of this paper. Here we show that, even simplified to the first point above, the total neutrino torque can still be reconstructed.

We construct an equivalent inertia $\I_R$ from $\T_c$, in order to maximum the cross-correlation coefficient $\mu$ between $\bs{\epsilon}\,\I_R\T_\nu$ and $\spin^I_{\nu 0}$.
In the primary coordinate of $\T_c$, $\T_c$ can be eigendecomposed as $\T_c=\sum_{i=1}^3\T^{\lambda_i}_c$ and we find $(\alpha_1,\alpha_2,\alpha_3)=(-0.7,0.7,0.08)$ (normalized such that $\sum \alpha_i^2=1$) and $\I_R\propto\sum_{i=1}^3\alpha_i\T^{\lambda_i}_c$
optimize $\mu$ to be 0.19.

\section{ERRORS}\label{app.error}
Consider $N$ halos with their unit spin vector randomly distributed on a 2-dimensional sphere, i.e., $|\spin|=1$ and $\left\langle \spin \right\rangle =\bs{0}$. Adding an additional vector $\epsilon\hat{\bs{x}}$ ($\epsilon\ll 1$) to $\spin$ and normalize, $\spin'=(\spin+\epsilon\hat{\bs{x}})/|\spin+\epsilon\hat{\bs{x}}|$, then project $\spin'$ onto $\hat{\bs{x}}$ and we get $p=\spin'\cdot\hat{\bs{x}}$. Since $\left\langle p \right\rangle=2\epsilon/3$ and $\sigma(p)=1/\sqrt{3N}$, an $n\sigma$ detection requires $N=3n^2/4\epsilon^2$.

When observational errors are considered, such as the scatter between halo and galaxy spins, we write the correlation between observed (galaxy) spin and underlying halo spin as $\left\langle \spin_{\rm obs} \cdot \spin' \right\rangle = \mu$, where $\mu\in [-1,1]$ is the the dot product of these two unit vectors.
    Projecting $\spin_{\rm obs}$ onto $\hat{\bs{x}}$ yields $\left\langle p_{\rm obs}\right\rangle
    =\left\langle\spin_{\rm obs}\cdot\hat{\bs{x}}\right\rangle
     = \mu\left\langle\spin'\cdot\hat{\bs{x}}\right\rangle
     =\mu\left\langle p\right\rangle
     =2\mu\epsilon/3$. 
     Here we assumed that the misalignment direction between halo and galaxy spins $\spin_{\rm obs}-\spin'$ is uncorrelated with the neutrino torque direction $\hat{\bs x}$.
     In this case $\sigma(p_{\rm obs})$ is still $1/\sqrt{3N}$, so an $n\sigma$ detection requires $N=3n^2/4|\mu|\epsilon^2$.


\bibliography{haoran_ref}

\begin{thebibliography}{10}
\expandafter\ifx\csname url\endcsname\relax
  \def\url#1{\texttt{#1}}\fi
\expandafter\ifx\csname urlprefix\endcsname\relax\def\urlprefix{URL }\fi
\providecommand{\bibinfo}[2]{#2}
\providecommand{\eprint}[2][]{\url{#2}}

\bibitem{2002PhRvL..89a1301A}
\bibinfo{author}{{Ahmad}, Q.~R.} \emph{et~al.}
\newblock \bibinfo{title}{{Direct Evidence for Neutrino Flavor Transformation
  from Neutral-Current Interactions in the Sudbury Neutrino Observatory}}.
\newblock \emph{\bibinfo{journal}{\prl}} \textbf{\bibinfo{volume}{89}},
  \bibinfo{pages}{011301} (\bibinfo{year}{2002}).

\bibitem{2014ChPhC..38i0001O}
\bibinfo{author}{{Olive}, K.~A.} \& \bibinfo{author}{{Particle Data Group}}.
\newblock \bibinfo{title}{{Review of Particle Physics}}.
\newblock \emph{\bibinfo{journal}{Chinese Physics C}}
  \textbf{\bibinfo{volume}{38}}, \bibinfo{pages}{090001}
  (\bibinfo{year}{2014}).

\bibitem{2018arXiv180706209P}
\bibinfo{author}{{Planck Collaboration}} \emph{et~al.}
\newblock \bibinfo{title}{{Planck 2018 results. VI. Cosmological parameters}}.
\newblock \emph{\bibinfo{journal}{arXiv e-prints}}
  \bibinfo{pages}{arXiv:1807.06209} (\bibinfo{year}{2018}).
\newblock \eprint{1807.06209}.

\bibitem{2011arXiv1110.3193L}
\bibinfo{author}{{Laureijs}, R.} \emph{et~al.}
\newblock \bibinfo{title}{{Euclid Definition Study Report}}.
\newblock \emph{\bibinfo{journal}{ArXiv e-prints}}  (\bibinfo{year}{2011}).
\newblock \eprint{1110.3193}.

\bibitem{2015AAS...22533605E}
\bibinfo{author}{{Eisenstein}, D.} \& \bibinfo{author}{{DESI Collaboration}}.
\newblock \bibinfo{title}{{The Dark Energy Spectroscopic Instrument (DESI):
  Science from the DESI Survey}}.
\newblock In \emph{\bibinfo{booktitle}{American Astronomical Society Meeting
  Abstracts}}, vol. \bibinfo{volume}{225} of \emph{\bibinfo{series}{American
  Astronomical Society Meeting Abstracts}}, \bibinfo{pages}{336.05}
  (\bibinfo{year}{2015}).

\bibitem{1998MNRAS.297L..71M}
\bibinfo{author}{{Mao}, S.}, \bibinfo{author}{{Mo}, H.~J.} \&
  \bibinfo{author}{{White}, S.~D.~M.}
\newblock \bibinfo{title}{{The evolution of galactic discs}}.
\newblock \emph{\bibinfo{journal}{\mnras}} \textbf{\bibinfo{volume}{297}},
  \bibinfo{pages}{L71--L75} (\bibinfo{year}{1998}).

\bibitem{2001ARA&A..39..137S}
\bibinfo{author}{{Sofue}, Y.} \& \bibinfo{author}{{Rubin}, V.}
\newblock \bibinfo{title}{{Rotation Curves of Spiral Galaxies}}.
\newblock \emph{\bibinfo{journal}{Annual Review of Astronomy and Astrophysics}}
  \textbf{\bibinfo{volume}{39}}, \bibinfo{pages}{137--174}
  (\bibinfo{year}{2001}).

\bibitem{2010MNRAS.404.1137B}
\bibinfo{author}{{Bett}, P.}, \bibinfo{author}{{Eke}, V.},
  \bibinfo{author}{{Frenk}, C.~S.}, \bibinfo{author}{{Jenkins}, A.} \&
  \bibinfo{author}{{Okamoto}, T.}
\newblock \bibinfo{title}{{The angular momentum of cold dark matter haloes with
  and without baryons}}.
\newblock \emph{\bibinfo{journal}{\mnras}} \textbf{\bibinfo{volume}{404}},
  \bibinfo{pages}{1137--1156} (\bibinfo{year}{2010}).
\newblock \eprint{0906.2785}.

\bibitem{2010MNRAS.405..274H}
\bibinfo{author}{{Hahn}, O.}, \bibinfo{author}{{Teyssier}, R.} \&
  \bibinfo{author}{{Carollo}, C.~M.}
\newblock \bibinfo{title}{{The large-scale orientations of disc galaxies}}.
\newblock \emph{\bibinfo{journal}{\mnras}} \textbf{\bibinfo{volume}{405}},
  \bibinfo{pages}{274--290} (\bibinfo{year}{2010}).
\newblock \eprint{1002.1964}.

\bibitem{2011MNRAS.415.2607D}
\bibinfo{author}{{Deason}, A.~J.} \emph{et~al.}
\newblock \bibinfo{title}{{Mismatch and misalignment: dark haloes and
  satellites of disc galaxies}}.
\newblock \emph{\bibinfo{journal}{\mnras}} \textbf{\bibinfo{volume}{415}},
  \bibinfo{pages}{2607--2625} (\bibinfo{year}{2011}).
\newblock \eprint{1101.0816}.

\bibitem{2015ApJ...812...29T}
\bibinfo{author}{{Teklu}, A.~F.} \emph{et~al.}
\newblock \bibinfo{title}{{Connecting Angular Momentum and Galactic Dynamics:
  The Complex Interplay between Spin, Mass, and Morphology}}.
\newblock \emph{\bibinfo{journal}{\apj}} \textbf{\bibinfo{volume}{812}},
  \bibinfo{pages}{29} (\bibinfo{year}{2015}).
\newblock \eprint{1503.03501}.

\bibitem{2018arXiv180407306J}
\bibinfo{author}{{Jiang}, F.} \emph{et~al.}
\newblock \bibinfo{title}{{Is the dark-matter halo spin a predictor of galaxy
  spin and size?}}
\newblock \emph{\bibinfo{journal}{ArXiv e-prints}}  (\bibinfo{year}{2018}).
\newblock \eprint{1804.07306}.

\bibitem{1984ApJ...286...38W}
\bibinfo{author}{{White}, S.~D.~M.}
\newblock \bibinfo{title}{{Angular momentum growth in protogalaxies}}.
\newblock \emph{\bibinfo{journal}{\apj}} \textbf{\bibinfo{volume}{286}},
  \bibinfo{pages}{38--41} (\bibinfo{year}{1984}).

\bibitem{2002MNRAS.332..325P}
\bibinfo{author}{{Porciani}, C.}, \bibinfo{author}{{Dekel}, A.} \&
  \bibinfo{author}{{Hoffman}, Y.}
\newblock \bibinfo{title}{{Testing tidal-torque theory - I. Spin amplitude and
  direction}}.
\newblock \emph{\bibinfo{journal}{\mnras}} \textbf{\bibinfo{volume}{332}},
  \bibinfo{pages}{325--338} (\bibinfo{year}{2002}).
\newblock \eprint{astro-ph/0105123}.

\bibitem{2015PhRvD..92b3502I}
\bibinfo{author}{{Inman}, D.} \emph{et~al.}
\newblock \bibinfo{title}{{Precision reconstruction of the cold dark
  matter-neutrino relative velocity from N -body simulations}}.
\newblock \emph{\bibinfo{journal}{\prd}} \textbf{\bibinfo{volume}{92}},
  \bibinfo{pages}{023502} (\bibinfo{year}{2015}).
\newblock \eprint{1503.07480}.

\bibitem{2017ApJ...847..110Y}
\bibinfo{author}{{Yu}, Y.}, \bibinfo{author}{{Zhu}, H.-M.} \&
  \bibinfo{author}{{Pen}, U.-L.}
\newblock \bibinfo{title}{{Halo Nonlinear Reconstruction}}.
\newblock \emph{\bibinfo{journal}{\apj}} \textbf{\bibinfo{volume}{847}},
  \bibinfo{pages}{110} (\bibinfo{year}{2017}).

\bibitem{2000ApJ...532L...5L}
\bibinfo{author}{{Lee}, J.} \& \bibinfo{author}{{Pen}, U.-L.}
\newblock \bibinfo{title}{{Cosmic Shear from Galaxy Spins}}.
\newblock \emph{\bibinfo{journal}{\apj}} \textbf{\bibinfo{volume}{532}},
  \bibinfo{pages}{L5--L8} (\bibinfo{year}{2000}).
\newblock \eprint{astro-ph/9911328}.

\bibitem{2001ApJ...555..106L}
\bibinfo{author}{{Lee}, J.} \& \bibinfo{author}{{Pen}, U.-L.}
\newblock \bibinfo{title}{{Galaxy Spin Statistics and Spin-Density
  Correlation}}.
\newblock \emph{\bibinfo{journal}{\apj}} \textbf{\bibinfo{volume}{555}},
  \bibinfo{pages}{106--124} (\bibinfo{year}{2001}).
\newblock \eprint{astro-ph/0008135}.

\bibitem{2017PhRvD..96l3502Z}
\bibinfo{author}{{Zhu}, H.-M.}, \bibinfo{author}{{Yu}, Y.},
  \bibinfo{author}{{Pen}, U.-L.}, \bibinfo{author}{{Chen}, X.} \&
  \bibinfo{author}{{Yu}, H.-R.}
\newblock \bibinfo{title}{{Nonlinear reconstruction}}.
\newblock \emph{\bibinfo{journal}{\prd}} \textbf{\bibinfo{volume}{96}},
  \bibinfo{pages}{123502} (\bibinfo{year}{2017}).

\bibitem{2018arXiv180706381W}
\bibinfo{author}{{Wang}, X.} \& \bibinfo{author}{{Pen}, U.-L.}
\newblock \bibinfo{title}{{Understanding the Reconstruction of the Biased
  Tracer}}.
\newblock \emph{\bibinfo{journal}{arXiv e-prints}}
  \bibinfo{pages}{arXiv:1807.06381} (\bibinfo{year}{2018}).
\newblock \eprint{1807.06381}.

\bibitem{2017MNRAS.469.1968P}
\bibinfo{author}{{Pan}, Q.}, \bibinfo{author}{{Pen}, U.-L.},
  \bibinfo{author}{{Inman}, D.} \& \bibinfo{author}{{Yu}, H.-R.}
\newblock \bibinfo{title}{{Increasing Fisher information by Potential Isobaric
  Reconstruction}}.
\newblock \emph{\bibinfo{journal}{\mnras}} \textbf{\bibinfo{volume}{469}},
  \bibinfo{pages}{1968--1973} (\bibinfo{year}{2017}).
\newblock \eprint{1611.10013}.

\bibitem{2017PhRvD..95d3501Y}
\bibinfo{author}{{Yu}, H.-R.}, \bibinfo{author}{{Pen}, U.-L.} \&
  \bibinfo{author}{{Zhu}, H.-M.}
\newblock \bibinfo{title}{{Nonlinear E -mode clustering in Lagrangian space}}.
\newblock \emph{\bibinfo{journal}{\prd}} \textbf{\bibinfo{volume}{95}},
  \bibinfo{pages}{043501} (\bibinfo{year}{2017}).
\newblock \eprint{1610.07112}.

\bibitem{2014ApJ...794...94W}
\bibinfo{author}{{Wang}, H.}, \bibinfo{author}{{Mo}, H.~J.},
  \bibinfo{author}{{Yang}, X.}, \bibinfo{author}{{Jing}, Y.~P.} \&
  \bibinfo{author}{{Lin}, W.~P.}
\newblock \bibinfo{title}{{ELUCID{\textemdash}Exploring the Local Universe with
  the Reconstructed Initial Density Field. I. Hamiltonian Markov Chain Monte
  Carlo Method with Particle Mesh Dynamics}}.
\newblock \emph{\bibinfo{journal}{\apj}} \textbf{\bibinfo{volume}{794}},
  \bibinfo{pages}{94} (\bibinfo{year}{2014}).
\newblock \eprint{1407.3451}.

\bibitem{2018ApJS..237...24Y}
\bibinfo{author}{{Yu}, H.-R.}, \bibinfo{author}{{Pen}, U.-L.} \&
  \bibinfo{author}{{Wang}, X.}
\newblock \bibinfo{title}{{CUBE: An Information-optimized Parallel Cosmological
  N-body Algorithm}}.
\newblock \emph{\bibinfo{journal}{The Astrophysical Journal Supplement Series}}
  \textbf{\bibinfo{volume}{237}}, \bibinfo{pages}{24} (\bibinfo{year}{2018}).

\bibitem{2012MNRAS.420.2551B}
\bibinfo{author}{{Bird}, S.}, \bibinfo{author}{{Viel}, M.} \&
  \bibinfo{author}{{Haehnelt}, M.~G.}
\newblock \bibinfo{title}{{Massive neutrinos and the non-linear matter power
  spectrum}}.
\newblock \emph{\bibinfo{journal}{\mnras}} \textbf{\bibinfo{volume}{420}},
  \bibinfo{pages}{2551--2561} (\bibinfo{year}{2012}).
\newblock \eprint{1109.4416}.

\bibitem{2019arXiv190401029Y}
\bibinfo{author}{{Yu}, H.-R.} \emph{et~al.}
\newblock \bibinfo{title}{{Probing primordial chirality with galaxy spins}}.
\newblock \emph{\bibinfo{journal}{arXiv e-prints}}  (\bibinfo{year}{2019}).
\newblock \eprint{1904.01029}.

\bibitem{2013JCAP...01..026A}
\bibinfo{author}{{Audren}, B.}, \bibinfo{author}{{Lesgourgues}, J.},
  \bibinfo{author}{{Bird}, S.}, \bibinfo{author}{{Haehnelt}, M.~G.} \&
  \bibinfo{author}{{Viel}, M.}
\newblock \bibinfo{title}{{Neutrino masses and cosmological parameters from a
  Euclid-like survey: Markov Chain Monte Carlo forecasts including theoretical
  errors}}.
\newblock \emph{\bibinfo{journal}{Journal of Cosmology and Astro-Particle
  Physics}} \textbf{\bibinfo{volume}{2013}}, \bibinfo{pages}{026}
  (\bibinfo{year}{2013}).
\newblock \eprint{1210.2194}.

\bibitem{2017RAA....17...85E}
\bibinfo{author}{{Emberson}, J.~D.} \emph{et~al.}
\newblock \bibinfo{title}{{Cosmological neutrino simulations at extreme
  scale}}.
\newblock \emph{\bibinfo{journal}{Research in Astronomy and Astrophysics}}
  \textbf{\bibinfo{volume}{17}}, \bibinfo{pages}{085} (\bibinfo{year}{2017}).
\newblock \eprint{1611.01545}.

\bibitem{2006astro.ph..6104P}
\bibinfo{author}{{Peterson}, J.~B.}, \bibinfo{author}{{Bandura}, K.} \&
  \bibinfo{author}{{Pen}, U.~L.}
\newblock \bibinfo{title}{{The Hubble Sphere Hydrogen Survey}}.
\newblock \emph{\bibinfo{journal}{ArXiv e-prints}}
  \bibinfo{pages}{astro--ph/0606104} (\bibinfo{year}{2006}).
\newblock \eprint{astro-ph/0606104}.

\bibitem{2018arXiv180207216D}
\bibinfo{author}{{Dawson}, K.} \emph{et~al.}
\newblock \bibinfo{title}{{Cosmic Visions Dark Energy: Small Projects
  Portfolio}}.
\newblock \emph{\bibinfo{journal}{ArXiv e-prints}}
  \bibinfo{pages}{arXiv:1802.07216} (\bibinfo{year}{2018}).
\newblock \eprint{1802.07216}.

\bibitem{2004MNRAS.350.1210Z}
\bibinfo{author}{{Zwaan}, M.~A.} \emph{et~al.}
\newblock \bibinfo{title}{{The HIPASS catalogue - II. Completeness, reliability
  and parameter accuracy}}.
\newblock \emph{\bibinfo{journal}{\mnras}} \textbf{\bibinfo{volume}{350}},
  \bibinfo{pages}{1210--1219} (\bibinfo{year}{2004}).
\newblock \eprint{astro-ph/0406380}.

\bibitem{2016PhRvD..93k3002D}
\bibinfo{author}{{Dvali}, G.} \& \bibinfo{author}{{Funcke}, L.}
\newblock \bibinfo{title}{{Small neutrino masses from gravitational
  {\ensuremath{\theta}} -term}}.
\newblock \emph{\bibinfo{journal}{\prd}} \textbf{\bibinfo{volume}{93}},
  \bibinfo{pages}{113002} (\bibinfo{year}{2016}).

\bibitem{2014PhRvL.113m1301Z}
\bibinfo{author}{{Zhu}, H.-M.}, \bibinfo{author}{{Pen}, U.-L.},
  \bibinfo{author}{{Chen}, X.}, \bibinfo{author}{{Inman}, D.} \&
  \bibinfo{author}{{Yu}, Y.}
\newblock \bibinfo{title}{{Measurement of Neutrino Masses from Relative
  Velocities}}.
\newblock \emph{\bibinfo{journal}{\prl}} \textbf{\bibinfo{volume}{113}},
  \bibinfo{pages}{131301} (\bibinfo{year}{2014}).
\newblock \eprint{1311.3422}.

\bibitem{2016PhRvL.116n1301Z}
\bibinfo{author}{{Zhu}, H.-M.}, \bibinfo{author}{{Pen}, U.-L.},
  \bibinfo{author}{{Chen}, X.} \& \bibinfo{author}{{Inman}, D.}
\newblock \bibinfo{title}{{Probing Neutrino Hierarchy and Chirality via
  Wakes}}.
\newblock \emph{\bibinfo{journal}{\prl}} \textbf{\bibinfo{volume}{116}},
  \bibinfo{pages}{141301} (\bibinfo{year}{2016}).
\newblock \eprint{1412.1660}.

\bibitem{2017NatAs...1E.143Y}
\bibinfo{author}{{Yu}, H.-R.} \emph{et~al.}
\newblock \bibinfo{title}{{Differential neutrino condensation onto cosmic
  structure}}.
\newblock \emph{\bibinfo{journal}{Nature Astronomy}}
  \textbf{\bibinfo{volume}{1}}, \bibinfo{pages}{0143} (\bibinfo{year}{2017}).

\bibitem{2017PhRvD..95h3518I}
\bibinfo{author}{{Inman}, D.} \emph{et~al.}
\newblock \bibinfo{title}{{Simulating the cold dark matter-neutrino dipole with
  TianNu}}.
\newblock \emph{\bibinfo{journal}{\prd}} \textbf{\bibinfo{volume}{95}},
  \bibinfo{pages}{083518} (\bibinfo{year}{2017}).

\bibitem{2019PhRvL.122d1302C}
\bibinfo{author}{{Chiang}, C.-T.}, \bibinfo{author}{{LoVerde}, M.} \&
  \bibinfo{author}{{Villaescusa-Navarro}, F.}
\newblock \bibinfo{title}{{First Detection of Scale-Dependent Linear Halo Bias
  in N -Body Simulations with Massive Neutrinos}}.
\newblock \emph{\bibinfo{journal}{\prl}} \textbf{\bibinfo{volume}{122}},
  \bibinfo{pages}{041302} (\bibinfo{year}{2019}).
\newblock \eprint{1811.12412}.

\end{thebibliography}

\end{document}